# Segregation growth and self-organization of ordered S atomic superlattices confined at interface between graphene and substrates

Dong-Lin Ma[§], Zhong-Qiu Fu[§], Ke-Ke Bai, Jia-Bin Qiao, Chao Yan, Yu Zhang, Jing-Yi Hu, Qian Xiao, Xin-Rui Mao, and Lin He*

**Ordered atomic-scale superlattices on surface hold great interest both for basic science and for potential applications in advanced technology. However, controlled fabrication of superlattices down to atomic scale has proven exceptionally challenging. Here we demonstrate the segregation-growth and self-organization of ordered S atomic superlattices confined at the interface between graphene and S-rich Cu substrates. Scanning tunneling microscope (STM) studies show that, by finely controlling the growth temperature, we obtain well-ordered S (sub)nanometer-cluster superlattice and monoatomic superlattices with various periods at the interface. These atomic superlattices are stable in atmospheric environment and robust even after high-temperature annealing (~ 350 ℃). Our experiments demonstrate that the S monoatomic superlattice can drive graphene into the electronic Kekulé distortion phase when the period of the ordered S adatoms is commensurate with graphene lattice. Our results not only open a road to realize atomic-scale superlattices at interfaces, but also provide a new route to realize exotic electronic states in graphene.**

Center for Advanced Quantum Studies, Department of Physics, Beijing Normal University, Beijing, 100875, People's Republic of China

[§]These authors contributed equally to this work.

Correspondence and requests for materials should be addressed to L.H. (e-mail: helin@bnu.edu.cn).

Creation and control of periodic atomic-scale structures to build nanodevices and nanosystems meet the emerging need of nanoelectronics and data storage[1–3]. Thanks to the invention of scanning tunneling microscope (STM)[4–6], it becomes possible to fabricate arbitrary atomic superlattices on metal surfaces by manipulation of STM tip[7–12]. However, the drawback of tip manipulation is obvious: the atomic-scale precision is almost impossible to realize on a large scale. Another promising route is self-organized growth of periodic adatoms on designer surfaces[13–18], which has advantages in fabrication of large-area atomic superlattices with tunable size and periodicity[19]. However, there are still two severe difficulties that need to overcome for the application of the atomic superlattices. One is that the atomic superlattices are usually only stable at low temperature and become unstable at high temperature due to thermal disturbance. The other is that the atomic superlattices on surface are easy to be destroyed because of absorption of other different atoms/molecules. Therefore, most of these delicately-built periodic atomic structures were realized in cryogenic environment and ultrahigh vacuum (UHV), which in principle hinder their potential application.

In this work, we demonstrate the idea of self-organization of atomic superlattice at the interface between graphene and the supporting substrate, which naturally overcome the said two shortcomings of the atomic superlattices at the surface. Interface could be seen as a Z-confined two-dimensional space compared to open space of surface[20,21]. The confined space provides extra stability against thermal disturbance and protects the "trapped" atoms or molecules at the interface from contamination. Here we realize self-organizing ordered S atomic superlattice at interfaces between graphene and S-rich Cu substrate. The ordered S adatoms are introduced into the interfaces by temperature-dependent segregation process and are found to be quite stable in atmospheric environment and robust even after high-temperature annealing (~350 ℃).

**Results**

**A general synthetic method for self-organized growth of sulfur superlattices at interface between graphene and substrates.** Figure 1 schematically shows chemical vapor deposition (CVD)-based growth process of graphene monolayer on Cu substrates.

To introduce S adatoms confined at the interface between graphene and the substrates, we use S-rich Cu foils as the supporting substrates[16,22-24]. The S atoms segregate from the Cu foils during the growth process, which is confirmed by our X-ray photoelectron spectroscopy (XPS) measurement (Supplementary Fig. 1), and form ordered atomic superlattices at the interface, as demonstrated subsequently in Fig. 2 and Fig. 3. In our experiment, the S-rich Cu substrates were annealed at high temperature as the first step to activate the S atoms both on the metal surface and in the bulk. In the second step, carbon sources were introduced into the system for graphene growth, then the sample was slowly cooled to room temperature (~20 ℃/min). We notice that the growth temperature, i.e. segregation starting temperature, has great influence on self-organization process of the S atoms at the interface. Well-ordered S (sub)nanometer-cluster superlattice and monoatomic superlattices with various periods are observed at the interface with different segregation starting temperatures.

**Self-organization of S (sub)nanocluster superlattice at interface.** Figure 2 presents STM characterizations of the graphene monolayer on Cu substrate intercalated with S atoms grown at 1000 ℃ (see Method and Supplementary Fig. 2 for details of the growth). The as-grown sample was delivered into UHV chamber from atmosphere and then annealed at ~350 ℃ for 1 hour for degassing. All STM measurements were performed at 4 K unless otherwise noted. Large-area square-ordered superlattices of the intercalated S atoms over several hundreds of nanometers can be observed in our STM measurement. Figure 2a shows a representative STM image of the graphene monolayer on Cu substrate intercalated with square-ordered S superlattices. In the zoom-in image (white dashed box in Fig. 2a), typical honeycomb lattices of the topmost graphene sheet were observed with blurred undercover features (Fig. 2a inset). The Fourier transform (FT) of the image, as shown in Fig. 2b, clearly reflects the detailed periodicity of graphene lattice (the outer six points) and the square-like S (sub)nanocluster superlattice (the inner four strongest scattering peaks), as well as their spatial relationship. The unit cell of the superlattice is rectangle with anisotropic periodicities of ~1.2 nm and ~1.5 nm respectively. It is also worth to mention that the long-axis direction of the superlattices is almost in parallel with that of graphene (rotation angle is smaller than 4°), indicating that the graphene plays a role in the formation of the superlattice.

Figure 2c shows atomic-resolved STM image of the intercalated S (sub)nanometer clusters obtained with low tunneling bias. Obviously, the (sub)nanometer clusters are the ordered assembly of S adatoms in a special "dice-five" shape (red circles in white box of Fig. 2c), which is similar as the structure of S reconstruction on Cu(100)[22]. Such a result, along with the result obtained by our XPS measurement, demonstrates explicitly that the studied sample is the graphene monolayer on Cu substrate intercalated with S atoms. It is also noted that the apparent height of the S clusters is only ~30 pm, strongly indicating that the S atom are almost buried in the topmost atomic layer of Cu substrate (Supplementary Fig. 3). All above evidences show that segregated S adatoms spontaneously assemble themselves into identical (sub)nanometer clusters, and self-organize into large-area hierarchical superlattice at the interface between graphene and Cu substrate. Importantly, the obtained (sub)nanoclusters superlattice is quite robust and still stable after several cycles of air exposure and degas annealing (~350 ℃) (seen in Supplementary Fig. 4).

To further explore mechanism of the self-organization of S atoms at the interfaces, we carefully studied structures around boundaries of the superlattices. Figure 2d shows a typical STM image measured around the boundary of the superlattices (indicated by blue lines). On the left region of the boundary, we do not observe the S (sub)nanoclusters superlattice. Instead, a typical one-dimensional moiré pattern generated by the lattices of graphene and Cu(100) facet is clearly observed. Interestingly, the one-dimensional moiré pattern on the left region is in line with the S superlattice on the right region, as indicated by red dashed lines in Fig. 2d. This strongly indicates that the self-organization process of S adatoms at interfaces is modulated by moiré superlattice of graphene and underlying Cu facets. It should be noted that missing-row defects (white dashed oval area in Fig. 2d) and adding-row defects (rectangle area) were occasionally observed around the boundary, suggesting that other factors such as local concentrations of S atoms also affect the formation of the superlattices. In Fig. 2e, we schematically model the three-layer structure of graphene, S superlattice and Cu(100). The graphene and Cu(100) were stacked in alignment (or 90°) and the moiré pattern could be visualized through the model simulation as marked in Fig. 2e (white dashed rectangle). The short-axis periodicity of the moiré supercell (~1.28 nm) is in coincidence with that of the S (sub)nanoclusters superlattice (red dashed rectangle). The long-axis periodicity of the moiré supercell (~6.7 nm) is much larger than that of the S (sub)nanoclusters superlattice (~ 4 times), indicating that there is other effect that plays a role in stabilizing more condensed arrays along the long axis of the moiré pattern[13].

**Self-organization of S adatoms superlattice at interface.** Figure 3 presents the STM characterization of the graphene monolayer on Cu substrate intercalated with S atoms grown at 850 ℃, in which the self-organization of S adatoms at interface exhibits quite a distinct feature. Figure 3a show the main feature of the sample that a large amount of dispersed atomic protrusions with a density of $(5.5\pm0.2)\times10^{14}$ cm$^{-2}$ could be visualized on graphene lattices. These protrusions were attributed to individual S adatoms trapped at the interface. The S adatoms locally exhibit square or hexagonal order (inset of Fig. 3a). However, there is no long-range order in large area. Such a result is further confirmed by FT of the image (Fig. 3b), which shows scattering-ring features (red dashed rings) besides the set of clear reciprocal lattice of graphene (black circles). We define the average radius of the rings as $1/\bar{R}$, where $\bar{R}$ corresponds to the average pair distance of the S adatoms. Extracted from the Fig. 3b, the value of $\bar{R}$ is estimated to be $0.65\pm0.15$ nm. The relatively low growth (segregation) temperature may be the reason that limits the kinetic energy or time of the S adatoms to form long-range order in large area. This is especially true for areas with high coverage of S adatoms. For areas with low coverage of S adatoms, it is relatively easy to observe ordered S adatoms superlattices confined at the interface. As shown in Fig. 3c, large-area ordered S adatoms superlattices with a S density of $(2.6\pm0.2)\times10^{14}$ cm$^{-2}$ were clearly observed at the area. The FT of the image (Fig. 3d) exhibits the clear q-space patterns of both clear graphene lattice (black circles) and S adatoms superlattice (first-order points marked by red circles). According to our experimental result, as shown in Fig. 3a and Fig. 3c, the positions of S adatoms have strong correlation with the high-symmetry points of graphene lattice, i.e., most of which are right underneath carbon atoms. It indicates that the graphene may play a more notable role of atomic templates in the formation of the adatoms superlattices than that of the S (sub)nanoclusters superlattice.

**Realization of Kekulé distortion (KD) phase on graphene modulated by special-ordered S adatoms superlattice.** The interaction between graphene and self-organized superlattices is mutual. Theoretical works have proposed that specific-ordered exotic adatoms could tune the electronic properties of graphene into new quantum phases beyond the pristine semimetal state[25–33]. One good example of novel quantum phases in graphene is the Kekulé distortion (KD) phase[34–37]. It is predicted that exotic adatoms on graphene would scatter electrons between two non-equivalent Dirac cones (valleys) at the corners of Brillouin zone (BZ) (Fig. 4a). The so-called intervalley scattering generates the Friedel oscillations (FO)—the electronic ripples in total charge density—in the vicinities of adatoms. When the arrangement of the adatoms on graphene follow an $(\sqrt{3} \times \sqrt{3})R30°$ (R3) order (demonstrated with three different colors in inset of Fig. 4a), the induced FO of individual adatoms would be in a coherent phase and function like the RKKY-type interaction between adatoms. The coherence of FO would significantly strengthen LDOS oscillations and lead to the emergence of KD phase in graphene. Fig. 4b shows a special area where strong herringbone-like charge-density wave (CDW) morphologies were observed on the same sample shown in Fig. 3. A close-up image (white box in Fig. 4b) shows that the one-dimensional strips (red dashed lines marked in Fig. 4b) are likely to be the corrugation of underlying substrate (Fig. 4c) and clear honeycomb lattices of graphene emerge over the whole area (white hexagons in Fig. 4c). Another important feature is that hexagonal contrasts of graphene lattice show a typical R3-ordered atomic features (red hexagons in Fig. 4c), which is exactly alike to the predicted KD phase. The simultaneously-acquired dI/dV mapping more clearly presents the R3-ordered nature of the CDW state (Fig. 4d). Above results evidence that a typical KD phase was observed and the stripped underlay may be the origin of the special symmetry-broken phase of graphene.

To elucidate the spatial configuration of graphene and substrate, FT of Fig. 4c is shown in Fig. 4e. Three sets of ordered patterns are marked in the image: the outer one corresponds to graphene lattice ($Q_g$, black circles); the intermediate pattern corresponds to the typical R3 superlattice which is assigned to KD phase ($Q_k$, blue circles); the rest inner scattering peaks (red circles) can be seen as a largely stretched hexagonal pattern of underlying "substrate", which is much alike to the reciprocal pattern of ordered S adatoms superlattice demonstrated in Fig. 3d. Fig. 4f schematically pictures the atoms of the "substrate" (red balls) in real space superimposed on graphene lattice by extracting the details of the pattern from the FT. It is clearly demonstrated that the atoms of the "substrate" is commensurate with graphene lattice, which could be described by two unequal integer sums of graphene vectors $L_1 = \vec{m} - 5\vec{n}$ (~1.13 nm) and $L_2 = \vec{m} + \vec{n}$ (~0.43 nm). The unique periodicity strongly indicates that the "substrate" is not any stable facets of Cu but a similar S adatoms superlattice, which strongly modulates the electronic properties of graphene. By further coloring the graphene lattice into three-color Kekulé texture (Fig. 4f upper right), it is surprisingly to find that the ordered adatoms sit in the same-color grid, which follow a hidden Kekulé order. We employed a phenomenological model of KD phase to further confirm the proposed configuration (left panel of Fig. 4g). A simulated STM image that is well matched to the experimental topography in Fig. 4b and Fig. 4c was created (right panel of Fig. 4g), suggesting of the validity of the proposed configuration of graphene and adatoms superlattice. This well explains the origin of such strong KD phase in this area and different morphologies of graphene with adatoms superlattice. It is also worth to mention that in this case the symmetry of graphene may take dominated roles in the formation of ordered adatoms superlattices at interface since the adatoms superlattice to a large extent follow the order of graphene lattice. As theory predicted, the KD phase possesses a gapped band structure which depends on strength of intervalley scattering induced by adatoms. As characterized by our STS measurements (Fig. 4h), a series of spectra with different tunneling current setpoints show a shared feature that an asymmetric bandgap of 245 ±5 mV appear in graphene ranging from -100 mV (top of the valence band) to 145 mV (bottom of the conduction band). The origin of the gap in KD phase was attributed to inequivalent electrons hopping $t_1$ and $t_2$ between nearest sublattices (inset in Fig. 4h). From fits to theoretical simulation, a similar gap would open at Γ point considering ~5% difference between $t_1$ and $t_2$, creating massive Dirac fermions of $m_D$= 0.28 ±0.02 $m_e$ (see method for details of calculation.

**Discussion**

Our experiments describe a method to create ordered S (sub)nanoclusters superlattice and adatoms superlattice through self-organization at the interface of graphene and Cu substrate. Both the symmetry of graphene and Cu substrate have influenced the self-organization process leading to various self-organized nanostructures. The superlattice with specific periodicity in turn modulate the electronic properties of graphene into new quantum phases like KD phase. The reported self-organization of atomic superlattices at interfaces may be universal and could be extended to other systems, which may provide a new route to realize exotic electronic states in graphene and other two-dimensional materials.

**Methods**

**CVD preparation of graphene.** The 25-μm Cu foil was purchased from Alfa Aesar. Before growth, Cu foil was first electropolished at 1.5 V DC voltage for 60 min, using a mixture of phosphoric acid and ethylene glycol (volume ratio = 3:1) as the electrolyte. The pre-treated Cu foil was loaded into a 2-inch quartz tube of low-pressure chemical vapor deposition furnace for sample growth. The Cu foil was first heated from room temperature to 1,000 ℃ in 30 min and kept for another 30 min, with 50 sccm (standard cubic centimeter per minute) $H_2$ and 50 sccm Ar as carrier gas. In the second step, the furnace temperature was set to 1000 ℃ (850 ℃) and then the carbon source was introduced into the furnace to grow graphene for 10 min. After all growth, the sample was cooled down to room temperature slowly (~20 ℃/min).

**STM and STS measurements.** STM/STS characterizations were performed in ultrahigh vacuum scanning probe microscopes (USM-1500 and USM-1400) from UNISOKU. The STM tips were obtained by chemical etching from a wire of Pt/Ir (80/20%) alloys. Lateral dimensions observed in the STM images were calibrated using a standard graphene lattice and a Si (111)-(7×7) lattice and Ag (111) surface. The dI/dV measurements were taken with a standard lock-in technique by turning off the feedback circuit and using a 793-Hz 5mV A.C. modulation of the sample voltage.

**Theoretical studies for the Kekulé phase in graphene.** We have adopted the tight-binding model to study the electronic structure of Kekulé superlattice. In this model, the operators with the two-atom (A and B sublattices) base for intrinsic graphene lattice without any distortion have turned into three two-atom bases which form the six-atom Kekulé basis. The two-atom bases are shown in Supplementary Fig. 5a. The operators $a_{i,j}$ and $b_{i,j}$ are the operators for the $i$th ($i$=1, 2, 3) atoms of three A sublattice and the three B sublattice atoms in the $j$th ($j$=I, II, III, IV) Kekulé basis, respectively. In this notation, the tight-binding Hamiltonian in real space is given by

$$H = -\sum_j (t a_{1,I}^\dagger b_{1,I} + t a_{1,I}^\dagger b_{2,I} + t a_{1,I}^\dagger b_{3,I}$$

$$+ t a_{2,I}^\dagger b_{1,III} + t a_{2,I}^\dagger b_{2,I} + t a_{2,I}^\dagger b_{3,IV}$$

$$+ t a_{3,I}^\dagger b_{1,III} + t a_{3,I}^\dagger b_{2,II} + t a_{3,I}^\dagger b_{3,I} + H.C.) \quad (S1)$$

Here t is the hopping amplitude dependent on the altered effective nearest neighbor distances. Based on previous DFT calculations, t empirically ranges from 2.7 eV to 3.16 eV. For simplify, in our superlattice, the hopping energies have two distinct values defined as $t_1$ and $t_2$.

Analogous to intrinsic graphene, the Hamiltonian can be transformed by writing the operators in reciprocal space

$$H_k = -\sum_k (t_l a_{i,k}^\dagger b_{i',k} e^{i\boldsymbol{k}\cdot(\boldsymbol{R}_j - \boldsymbol{R}_{j'})} + H.C.) \quad (S2)$$

Where $\boldsymbol{R}_j - \boldsymbol{R}_{j'}$ is the effective distance between the corresponding Kekulé unit cells.

By diagonalizing Hamiltonian (S2), we obtain the low-energy electronic spectra (the 3$^{rd}$ and 4$^{th}$ sub-bands of six-band spectra) of the Kekulé superlattice (Supplementary Fig. 5b).

**Acknowledgements**

This work was supported by the National Natural Science Foundation of China (Grant Nos. 11674029, 11422430, 11374035), the National Basic Research Program of China (Grants Nos. 2014CB920903, 2013CBA01603), the program for New Century Excellent Talents in University of the Ministry of Education of China (Grant No. NCET-13-0054), the China Postdoctoral Science Foundation (No. 212400207). L.H. also acknowledges support from the National Program for Support of Top-notch Young Professionals and support from "the Fundamental Research Funds for the Central Universities".

**Author contributions**

D.L.M., Z.Q.F., K.K.B., C.Y., J.Y.H., Q.X. and X.R.M. synthesized the samples. D.L.M. and Z.Q.F. performed the STM experiments. D.L.M., Z.Q.F., J.B.Q. and Y.Z. analyzed the data. L.H. conceived and provided advice on the experiment and analysis. D.L.M., Z.Q.F. and L.H. wrote the paper. All authors participated in the data discussion.

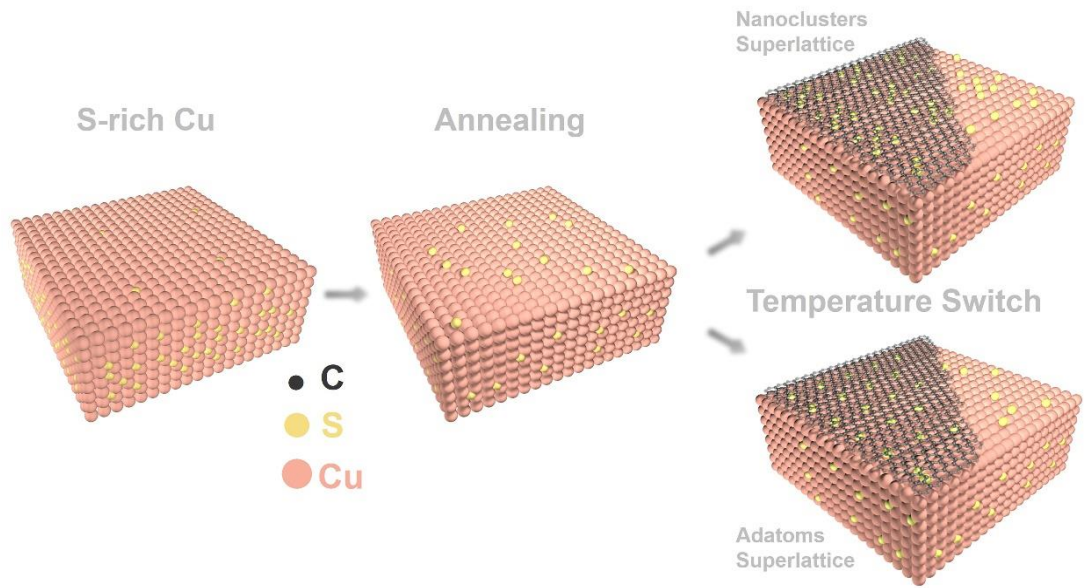

**Figure 1. A general synthetic method for self-organized growth of sulfur superlattices in between graphene and substrates.** In the first step, the growth substrates were annealed at high temperature to dissolve and activate S atoms in the bulk. In the following, carbon sources were introduced into the system for graphene growth, then the samples were slowly cooled to room temperature. During the slow cooling process, vast of S atoms begun to segregate from the bulk onto the metal surfaces due to the decrease of solubility. At this stage, the S adatoms had sufficient energy and time to rearrange beneath the as-grown graphene sheet to form kinds of ordered assemblies.

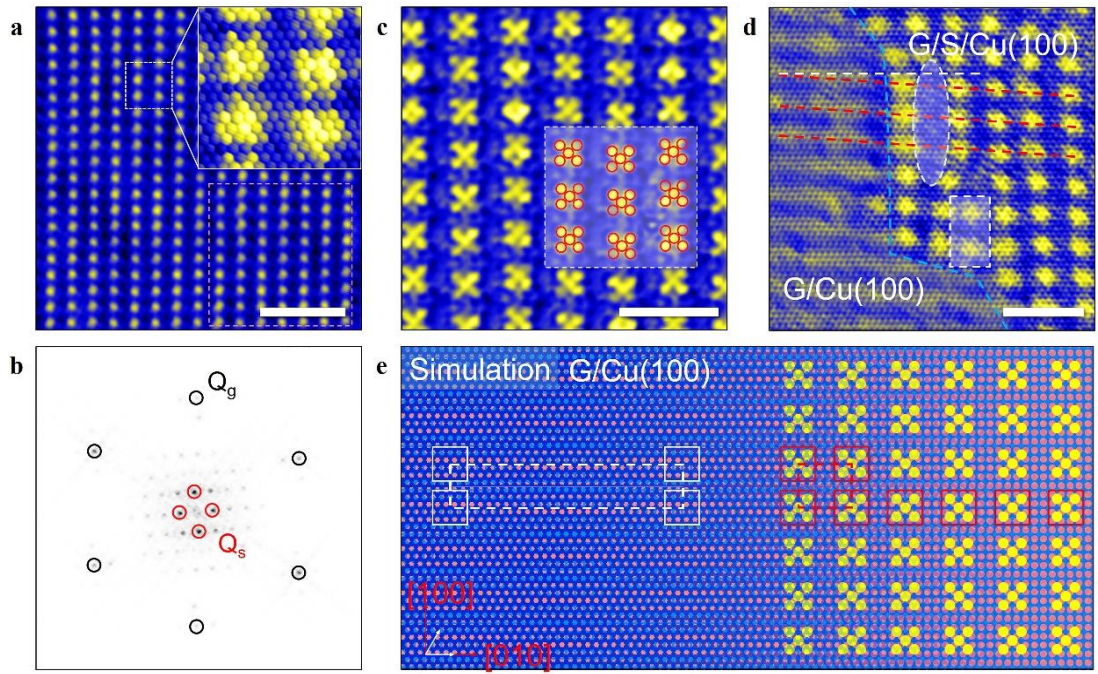

**Figure 2. Self-organization of S (sub)nanoclusters superlattice at interface.** (a) STM topography of large-area square-ordered superlattices ($V_b$ = 800 mV, I = 400 pA). Inset (white dashed box) shows zoom-in image, clearly exhibiting honeycomb lattices of graphene. (b) Fourier transform (FT) of the STM image in (a). (c) Atomic-resolved STM image of the superlattice under graphene ($V_b$ = 50 mV, I = 150 pA). The nanoclusters are the ordered assembly of adatoms in a special "dice-five" shape, as illustrated by red circles in white box. (d) A boundary of the (sub)nanoclusters superlattice ($V_b$ = 50 mV, I = 200 pA). One-dimensional moiré patterns are indicated by red dashed lines. The missing-row defects and adding-row defects of the superlattice around the boundary were marked in white dashed oval and rectangle, respectively. (e) Schematics of the three layer configuration of graphene, S superlattice and Cu(100) substrate. The moiré pattern generated by the misalignment of graphene and Cu(100) is illustrated by white dashed rectangle. The unit cell of S superlattice is illustrated by red dashed rectangle. The scale bars are 6 nm in (a), 3 nm in (b), and 3 nm in (c).

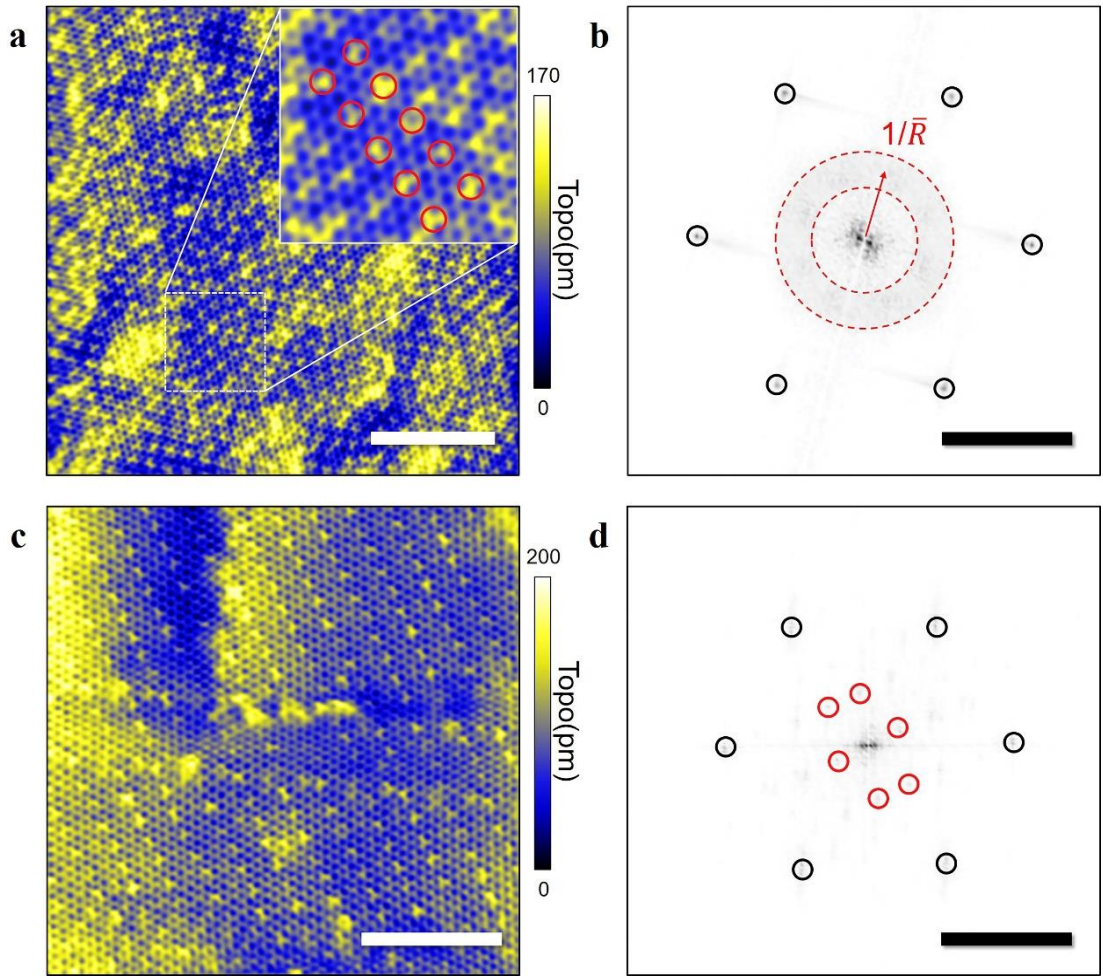

**Figure 3. Self-organization of S adatoms superlattice at interface.** (a) STM topography of large-area graphene lattices with a large amount of dispersed adatoms underneath graphene ($V_b$ = 200 mV, I = 250 pA). Inset shows that these S adatoms locally show a weak square order. (b) FT of the (a) showing scattering-ring feature. (c) Large-area ordered S adatoms superlattice with clear graphene lattice ($V_b$ = 650 mV, I = 250 pA). (d) Corresponding FT image of (c). The scale bars are 4 nm (a), 4 nm$^{-1}$ (b), 3 nm (c), and 4 nm$^{-1}$ (d).

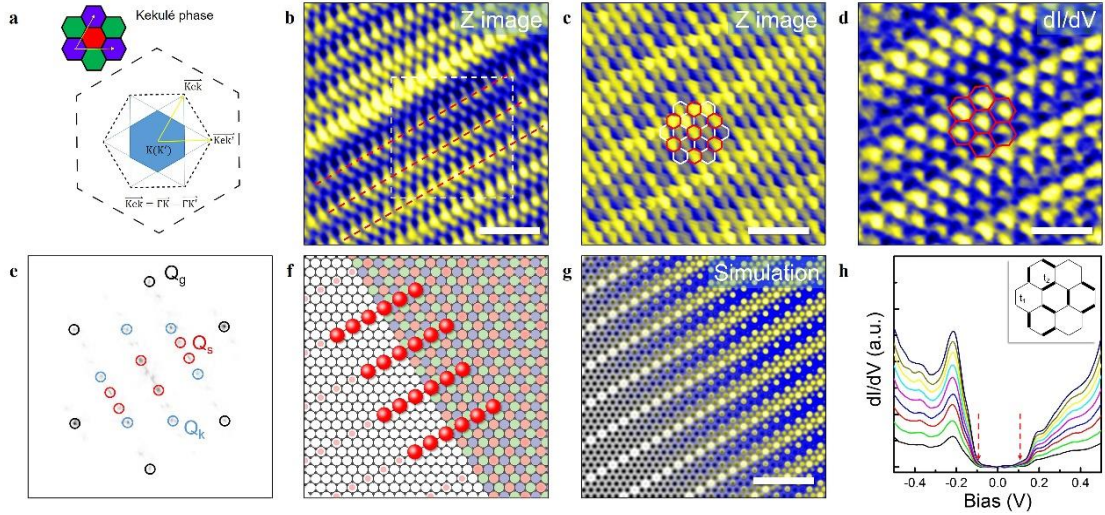

**Figure 4. Realization of Kekulé distortion (KD) phase on graphene modulated by special-ordered S adatoms superlattice.** (a) Brillouin zone and real-space schematics (upper left) of graphene in Kekulé distortion (KD). (b) STM topographic image of strong herringbone-like charge-density wave (CDW) morphologies on the sample of Fig. 3 ($V_b$ = 300 mV, I = 200 pA). (c) Low-bias image (white box area in (a)) showing honeycomb lattice of graphene and R3-ordered atomic features, as marked by white and red hexagons ($V_b$ = 30 mV, I = 200 pA). (d) The simultaneously-acquired dI/dV mapping ($V_b$ = 218 mV, I = 200 pA). (e) Fourier transform of (c). (f) Schematic illustrations of underneath adatoms superimposed on color-coded graphene lattice. (g) Phenomenological model of KD phase on the configuration of (f). (h) STS spectra on the KD phase with different tunneling current setpoints (from 200 to 1000 pA). A bandgap of ~245 mV was observed in this regions. Edges of valence band and conduction band are indicated by arrows. Inset shows honeycomb models for calculation, with inequivalent electrons hoppings $t_1$ and $t_2$ between nearest sublattices. All scale bars are 2 nm.